\documentclass[12pt]{iopart}
\usepackage{graphicx}
\usepackage{bm}

\newcommand{\tabref}[1]{Table~\ref{#1}}
\newcommand{\equref}[1]{eq.~(\ref{#1})}
\newcommand{\figref}[1]{Fig.~\ref{#1}}
\newcommand{\secref}[1]{Section~\ref{#1}}

\begin{document}

\title{Flagellar Dynamics of Chains of Active Janus Particles Fueled by an AC electric field}

\author{Daiki Nishiguchi$^{1}$\footnote{Current addresses: Service de Physique de l'Etat Condens\'e, CEA, CNRS, Universit\'e Paris-Saclay, CEA-Saclay, 91191 Gif-sur-Yvette, France; Pathogenesis of vascular infections unit, Institut Pasteur, 75015 Paris, France.}, Junichiro Iwasawa$^1$, Hong-Ren Jiang$^2$ \& Masaki Sano$^1$}

\address{$^1$ Department of Physics, The University of Tokyo, Hongo 7-3-1, Tokyo, 113-0033, Japan.}
\address{$^2$ Institute of Applied Mechanics, National Taiwan University, No.1, Sec. 4, Roosevelt Rd., Da'an Dist., Taipei City 106, Taiwan.}

\ead{sano@phys.s.u-tokyo.ac.jp}
\vspace{10pt}
\begin{indented}
\item[]August 2017
\end{indented}

\begin{abstract}
We study the active dynamics of self-propelled asymmetrical colloidal particles (Janus particles) fueled by an AC electric field. Both the speed and the direction of the self-propulsion and the strength of attractive interaction between the particles can be controlled by tuning the frequency of the applied electric field and the ion concentration of the solution. The strong attractive force at high ion concentration give rise to chain formation of the Janus particles, which can be explained by the quadrupolar charge distribution on the particles. The chain formation is observed irrespective of the direction of the self-propulsion of the particles.
When both the positions and the orientations of the heads of chains are fixed, they exhibit beating behavior reminiscent of eukaryotic flagella. The beating frequency of the chains of the Janus particles depends on the applied voltage and thus on the self-propulsive force. The scaling relation between the beating frequency and the self-propulsive force deviates from theoretical predictions made previously on active filaments. However, this discrepancy is resolved by assuming that the attractive interaction between the particles is mediated by the quadrupolar distribution of the induced charges, which gives indirect but convincing evidence on the mechanisms of the Janus particles. This signifies that the dependence between the propulsion mechanism and the interaction mechanism, which had been dismissed previously, can modify dispersion relations of beating behaviors. In addition, hydrodynamic interaction within the chain and its effect on propulsion speed are discussed. These provide new insights on active filaments such as optimal flagellar design for biological functions.
\end{abstract}


%
%
%
%
%

\section{Introduction}
Active matter systems---collections of active elements that transduce free energy into motion---display a vast variety of intriguing nonequilibrium phenomena.
Especially, collective motion of self-propelled particles interacting with each other has been intensively studied both theoretically and experimentally in the past few decades.
Flocking phases with true- or quasi-long-range orientational order has been studied theoretically and are now recognized as one of the most well-understood classes of collective motion \cite{Toner2005,Ramaswamy2010,Marchetti2013}.
The theoretical studies have predicted nontrivial properties such as giant density fluctuations in these globally ordered phases that contrast equilibrium systems, although never have such properties been experimentally observed until very recently \cite{Nishiguchi2017}. Another example of theoretically well-studied collective motion is ``active turbulence'' observed in dense suspensions of swimming or swarming bacteria \cite{Dombrowski2004,Sokolov2007,Sokolov2012,Wensink2012} and self-propelled colloids \cite{Nishiguchi2015}, and it has been to some extent successfully described by hydrodynamic equations \cite{Marenduzzo2007,Wensink2012,Dunkel2013njp,Dunkel2013prl} and kinetic theories \cite{Grossmann2014}.
Interacting self-propelled Janus particles, driven by diffusiophoresis in a near-critical mixture \cite{Buttinoni2013} or fueled by hydrogen peroxide \cite{Theurkauff2012,Palacci2013}, are reported to exhibit dynamical clusters or aggregates, where particles are constantly going in and out without forming strong connections between the particles.

Examples of active matter systems raised above are concerned about groups of individual active elements interacting with each other without any configurational coupling or topological connectivity. How do active systems under such constraints behave?
In recent works \cite{Kaiser2015,Laskar2015,Eisenstecken2016,Eisenstecken2017}, chains of self-propelled particles have been theoretically investigated not only due to purely physical interest on extensions of concepts in polymer science such as the polymer scaling theory but also because of the resemblance of such systems to biological ones. Eukaryotic flagella as seen in spermatozoa, algae, etc., for example, contain molecular motors as well as passive protein filaments. Internal stress generated by the molecular motors accumulates and eventually deforms the filaments, leading to their beating motion, which is utilized for swimming in the low Reynolds number world \cite{Purcell1976}. Such beating behavior has been realized in reconstituted systems on microtubules and molecular motors \cite{Sanchez2011}, and theoretical models have predicted some scaling relations among beating frequency, internal force, bending rigidity, etc \cite{Sekimoto1995,Julicher1995,Camalet1999}. Numerical studies have also pointed out that similar beating behavior might be observed in chains of connected self-propelled particles \cite{Jayaraman2012,Chelakkot2014}.
However, in spite of significant biological importance, to the best of our knowledge, the difficulty in connecting self-propelled particles on microscopic scales has hindered experimental exploration on the active dynamics of chains of self-propelled particles.

Here, by elucidating propulsion and interaction mechanisms of self-propelling Janus particles fueled by a vertical AC electric field that swim horizontally, we realized microscopic ``self-propelled chains'' composed of self-propelled colloidal particles with both tunable propulsive force and interactions [\figref{fig1}(b)]. We focus on their beating behavior observed when the fore-most particles of the chains are fixed on a substrate that resembles that of eukaryotic flagella. 
An intriguing property of chains of the active Janus particles is that the formation of chains is self-organized thus reconfigurable. Due to this property, both attractive force between particles and propulsive force are dependent on the applied vertical electric field. This may result in different scaling relations of dispersion relation of beating behavior from that of eukaryotic flagella.
Controllability of the propulsive force and the interactions enabled us to explore how beating behaviors in active systems depends on the strength of internal stress generation by each element. We address this question by experimentally measuring the dispersion relation of waves propagating on the chains.
Scaling relations obtained in our experiment deviates from the theoretical predictions made previously \cite{Sekimoto1995,Bourdieu1995,Chelakkot2014}, but we have resolved this discrepancy by taking into account the mechanisms of the propulsion and the interactions of the Janus particles. This not only supports the theoretical predictions but also accounts for the precise mechanisms of propulsion and interactions, especially the distribution of induced charges, of Janus particles fueled by an AC electric field. Our results also provide a new perspective on beating active filaments and their design principles.

\section{Experimental setup}
\begin{figure}[t]
\begin{center}
\includegraphics[width=138truemm]{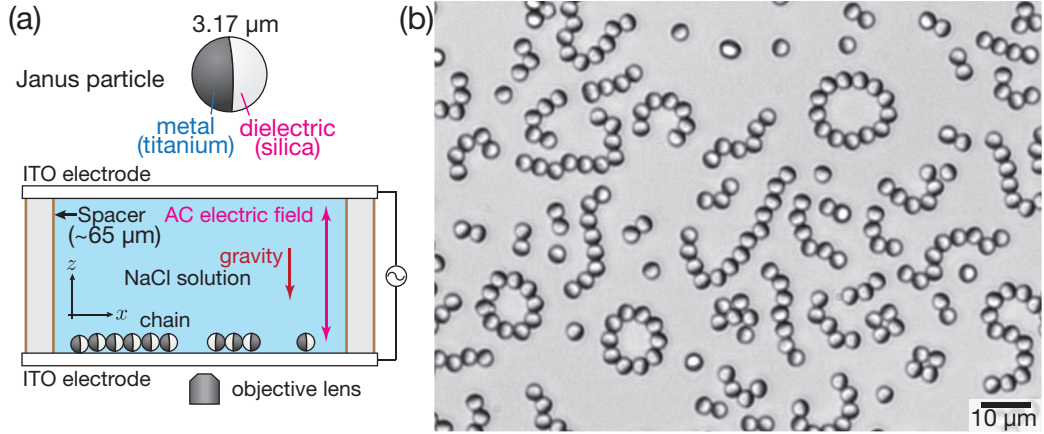}
\caption{
(a) Schematics of the Janus particle and the experimental setup. Suspension of Janus particles is sandwiched between two transparent electrodes.
(b) Experimental snapshot (1 MHz, 30 $\mathrm{V_{pp}}$, 0.1 mM). Attractive interaction creates chains and complexes of self-propelled Janus particles. Intensity is adjusted for visibility. See also Supplemental Movie 1.
\label{fig1}}
\end{center}
\end{figure}

We used metallodielectric Janus colloids as self-propelled particles, which have two distinct hemispheres with different properties [\figref{fig1}(a)]. 
The Janus particles were fabricated from purchased silica ($\mathrm{SiO_2}$) colloidal particles (diameter $3.17~\mathrm{\mu m}$, Bangs Laboratories, Inc., SS05N). We first deposited titanium by using electron-beam deposition (thickness: 35~nm), and then silica by thermal evaporation (thickness: 15~nm) on the silica colloids in order to ensure that the whole surface is covered with silica. 
The Janus particles were resuspended in NaCl solutions with concentrations ranging from 0.1 mM to 1.0 mM, and washed with sonication. This washing procedure was repeated 3 times after waiting for 30--60 minutes for sedimentation and replacing the NaCl solution. We used NaCl solution instead of deionized water, because the ion concentration in solutions increases over time due to dissolution of ions from the surfaces of experimental apparatus and gases in the air during preparation and experiments, resulting in unsteady drift in the electrical conductivity of the solution. The addition of NaCl decreases relative drift in ion concentration and enables reliable measurements.

The suspension was sandwiched between two transparent indium tin oxide (ITO) electrodes (Mitsuru Optical Co. Ltd., Japan) separated by stretched Parafilm (thickness: $\sim65~\mathrm{\mu m}$) [\figref{fig1}(a)]. These ITO electrodes were also coated with a 25-nm layer of silica by thermal evaporation beforehand in order to avoid adhesion of the Janus particles on the electrodes \cite{Yan2016}. 
Before turning on the electric field, we waited for a couple of minutes to ensure that all the Janus colloids were sedimented very close to the bottom electrode to form a quasi-two-dimensional layer. In addition to this, hydrodynamic torque exerted by induced-charge electro osmotic (ICEO) flows around the particles orients and thus always confines polarities of the Janus particles in the horizontal $xy$-plane upon the application of an electric field \cite{Kilic2011a}, ensuring two-dimensional motion of the particles \cite{Nishiguchi2015}.
Transmitted bright-field images of particles were captured through the bottom electrode by using an inverted microscope (Nikon ECLIPSE TE2000-U) with an objective lens (Nikon Plan Fluor ELWD, $40\times$, NA=0.60) and a CMOS camera (Baumer HXG40 or Baumer LXG80). The particles were illuminated by a halogen lamp with a green filter (GIF), and in this condition, the titanium sides and the silica sides of the particles look dark and white respectively [\figref{fig1}(b) and Supplemental Movie 1].

\section{Frequency dependence and chain formation}
\label{SecFormationOfChains}
The self-propulsion mechanism of Janus particles under an AC electric field is partially explained by induced-charge electro-osmosis (ICEO) and electrophoresis (ICEP) \cite{Squires2004,Squires2006,Gangwal2008}. When the electric field is turned on, surface charges of the particle induced by the electric field gathers counter ions in the bulk solution, which leads to the formation of the electric double layer (EDL). The counter ions in the EDL is driven by the electric field, forming electro-osmotic flows around the particles (ICEO). The difference in the physical properties of the hemispheres leads to asymmetric flows, resulting in self-propulsion perpendicular to the electric field (ICEP). 
However, this ICEP theory is unable to predict how the the Janus particles actually behave under an AC electric field with the non-zero frequency $f$.
The ICEP theory is indeed consistent with some experimental observations such as quadratic dependence of the self-propulsion velocity $v_0$ on the applied voltages $V$, $v_0\propto V^2$. However, the low frequency limit $f\to 0$ taken in the theoretical calculation has obscured some experimental facts such as velocity reversal and attractive interactions at high frequencies \cite{JiangUnpublished,Suzuki2011,SuzukiThesis,Yan2016}. Although the dielectric spectrum has been calculated in Refs.~\cite{JiangUnpublished,SuzukiThesis, Yan2016}, understanding the mechanism of the velocity reversal and the interaction between Janus particles at high frequency is far from satisfactory.

\begin{figure}[t]
\begin{center}
\includegraphics[width=138truemm]{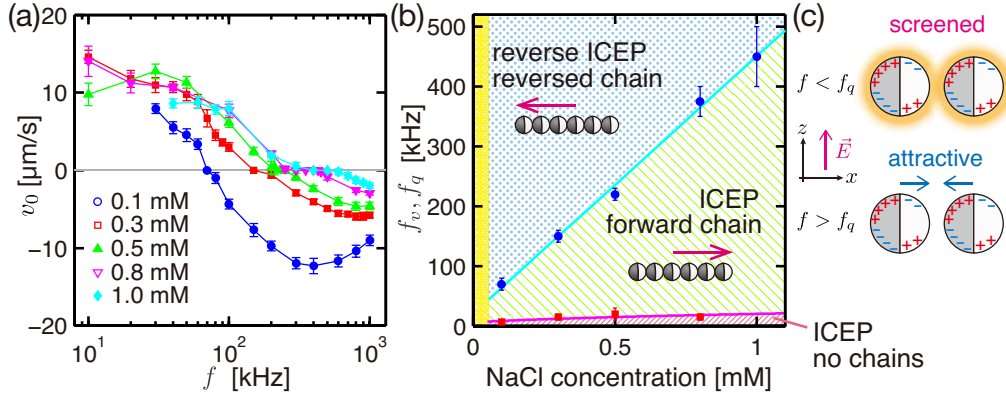}
\caption{
(a) Frequency dependence of self-propulsion velocity $v_0$ for different NaCl concentrations ($V=16$ $\mathrm{V_{pp}}$). Error bars: Standard deviations among the particles.
(b) The velocity reversal frequency $f_v$ (blue circles) and the interaction switching frequency $f_q$ (red squares) for different NaCl concentrations. Above $f_q$, the attractive interaction leads to chain formation. $f_q$ is visually estimated from the threshold frequency at which chains become dissociated as we lower $f$, and thus the values of $f_q$ shown here correspond to lower bounds of $f_q$ (see Supplemental Movie 3). A proportional relation between $f_v$ and the ion concentration is obtained. Solid lines: fitting curves with linear (cyan, $f_v$) and $f_q\propto\sqrt{c_\mathrm{NaCl}+c_0}$ where $c_\mathrm{NaCl}$ is the concentration of NaCl (magenta, $f_q$). Error bars for $f_v$: uncertainty of the zero-crossing points in (a). Error bars for $f_q$: uncertainty from the visual inspection. Yellow shaded region: no precise measurement due to large drift in ion concentration over time.
(c) Schematics of the quadrupolar distribution of the induced charge and the consequent attractive interactions. Top: Below the frequency $f_q$, the counter ions and the electric double layer (EDL) screen the quadropoles on the Janus particles (orange clouds) and thus the interaction is too week to form chains. Bottom: Above $f_q$, the EDL is thin enough for the attractive interaction between the quadrupoles to appear. 
\label{fig2}}
\end{center}
\end{figure}

Here, in addition to the frequency dependence of the Janus particles driven by ICEP studied previously \cite{Suzuki2011,SuzukiThesis,Yan2016}, we introduced the ion concentration as a new control parameter in order to explore a broader parameter space for better understanding of the mechanism.
\figref{fig2}(a) shows the frequency dependence of the propulsion velocity $v_0$ of isolated Janus particles from 10 kHz to 1 MHz for different NaCl concentrations. At low frequencies ($\sim10^2$ kHz), particles moved towards their silica hemispheres, which is consistent with the ICEP theory \cite{Squires2006}. However, in the high frequency region above a threshold frequency $f_v$ ($f_v \sim70$ kHz for 0.1 mM), particles moving toward their titanium hemispheres were observed (reverse ICEP: rICEP).
This velocity reversal of Janus particles has been reported in the previous works \cite{Suzuki2011,SuzukiThesis,Yan2016}, but still lacks a theoretical explanation.
We obtained the linear frequency dependence between $f_v$ and the ion concentration [\figref{fig2}(b)], which may give a clue for future theoretical development.

We observed chain formation of the Janus particles not only in the rICEP regime around $f\sim1$ MHz as reported previously \cite{JiangUnpublished,Suzuki2011,Yan2016} but also in the usual ICEP regime at high NaCl concentration around $\sim1$ mM, the parameters unexplored previously (see \figref{fig3} and Supplemental Movies 1 and 2).
This signifies that the two transitions of the self-propulsion direction and the interactions have different characteristic frequencies $f_v$ and $f_q$ respectively and thus have different mechanisms that were thought to be the same previously \cite{JiangUnpublished,Suzuki2011,Yan2016}. In order to estimate $f_q$, we firstly increase the frequency of the electric field $f$ to a higher value for Janus particles to form chains spontaneously, and then we suddenly jump the frequency toward the target frequency to inspect whether the chains dissociate or not at the targeted voltage [\figref{fig2}(b)]. Because this value of $f_q$ is estimated not from formation but from dissociation of chains, this estimate gives a lower bound of $f$ for sustaining chain structure, which does not mean that Janus particles form chains as soon as we increase $f$ above $f_q$. 

The formation of chains can be explained by a quadrupole-like distribution of induced charges on a Janus particle. The charge distribution on the sphere with radius $a$ induced by the electric filed $\bm{E}_0$ can be obtained by solving the Laplace equation around the sphere with the boundary conditions; ${\displaystyle (\epsilon_l \bm{E}_l - \epsilon_p \bm{E}_p) \cdot \bm{e}_{\bm{r}}|_{r=a} = \sigma_f}$ and  
${\displaystyle (\gamma_l \bm{E}_l - \gamma_p \bm{E}_p)\cdot \bm{e}_{\bm{r}}|_{r=a} = -(\partial \sigma_f/\partial t)}$, where $\epsilon_i$ and $\gamma_i$ refer to the dielectric permittivity and the conductivity, $i=l,\; p$ represents the liquid and the particle respectively, and $\sigma_f$ is the free charge on the surface.
For a conducting spherical particle in a conducting fluid such as a NaCl solution, the effective induced dipole, $\bm{P}_\mathrm{eff}$, is given by \cite{Jones1979,Pannacci2007}
\begin{eqnarray}
\bm{P}_\mathrm{eff}&=&4\pi \epsilon_l a^3 \frac{\gamma_p - \gamma_l}{\gamma_p + 2\gamma_l}\bm{E}_0 \quad \mbox{for} \quad f \ll f_M, \quad \mbox{(conducting regime)} \\
\bm{P}_\mathrm{eff}&=&4\pi \epsilon_l a^3 \frac{\epsilon_p - \epsilon_l}{\epsilon_p + 2\epsilon_l}\bm{E}_0 \quad \mbox{for} \quad f \gg f_M \quad \mbox{(dielectric regime)}
\end{eqnarray}
where $f_M=\frac{1}{2\pi} \frac{\epsilon_p + 2 \epsilon_l}{\gamma_p + 2 \gamma_l}$ is the Maxwell-Wagner frequency which is the order of $10^{15}$ Hz for a metallic particle and $10^5$ Hz for a dielectric particle immersed in water or dilute ionic solutions.
We consider the Janus particle as a superposition of two different spherical particles, one hemisphere represents the property of a metallic sphere and the other hemisphere represents that of a dielectric material. For the metallic side, since $\gamma_l < \gamma_p$ always holds, $\bm{P}_\mathrm{eff}$ is parallel to $\bm{E}_0$ for the whole frequency range in the experiment. 
Induced surface charges will be neutralized by counter ions in the EDL.
While for the dielectric side, $\gamma_l > \gamma_p$ holds for diluted NaCl solutions. Therefore the induced dipole is anti-parallel to $\bm{E}_0$ on the dielectric side. This is true even for frequencies higher than $f_M$, because the dielectric constant of water is much higher than that of usual dielectric materials ($\epsilon_p < \epsilon_l$). The counter ions near the surface of dielectric materials have opposite signs (see Fig.2(c)). 
The electric field drives charges in the EDL from the poles of the particle to the equator on the metal side, while on the dielectric side the surface slip osmotic flow is from the equator to the poles. This expected flow pattern exactly coincides with the observation of flow pattern visualized for Janus particles under an AC electric field at 1~kHz by Peng {\it et al} \cite{Peng2014}\footnote{Directionality of the induced dipole can be also tested by checking the Quincke effect and consequent rotation of particles \cite{Jakli2008,Bricard2013}. If the induced dipole is antiparallel to the applied electric field, the particle will rotate due to the instability of unstable configuration above a certain critical electric field. The same thing holds for an AC electric field at least for frequencies lower than the the RC frequency. We observed that silica particles in 0.1mM NaCl solution exhibits Quincke rotation under an AC electric field at 1~kHz. That implies the induced dipole on the dielectric side is antiparallel to $\bm{E}_0$. A logical consequence is that the directionality of the induced dipole on the metallic hemisphere of a Janus particle is parallel to $\bm{E}_0$, because the Janus particle would rotate due to the Quincke effect if the induced dipoles on both sides of the Janus particle were antiparallel to $\bm{E}_0$. As a conclusion, the surface charge distribution on Janus particles can be justified as a quadrupole-like distribution as shown in \figref{fig2}(c).}.
 
Therefore, two Janus particles with quadrupolar charge distributions sitting on the plane attracts each other in principle (see Fig.2(c)). However, the electrostatic interaction between the two particles is screened by the EDL surrounding the Janus particles. As the frequency of the applied AC electric field is increased beyond inverse of the so-called RC time, the electric screening effect will be substantially reduced. The RC time
$\tau_{\rm RC}$ is given by $\tau_{\rm RC}={a \epsilon}/{\gamma \lambda_{\rm{D}}}$ using Debye-H\"{u}ckel approximation where $a$ is the radius of the sphere, $\gamma$ is the conductivity of the solution and $\lambda_{\rm D}$ is the Debye length.  $\tau_{\rm RC}$ corresponds to the charging time of the EDL capacitance, $C=4\pi a^2 \epsilon/\lambda_{\rm D}$ through the resistance of the bulk electrolyte, $R = 1/4\pi \gamma a$ \cite{GarciaSanchez2012}. 
The formation of the EDL cannot keep up with the electric field at high AC frequencies and consequently the screening of the surface charges becomes weaker.  In our system, the RC time for Janus particles suspended in a 0.1 mM NaCl solution is about $1/\tau_{\rm RC}\sim 10$ kHz. 
Above such a frequency, two Janus particles can come close to each other where quadrupole-quadrupole interaction exhibits attractive interaction. 
Therefore, the characteristic frequency for chain formation $f_q$ can be explained by the RC time as $f_q\sim1/\tau_{\rm RC}$. The increase of $f_q$ with increasing the concentration of NaCl can be explained by the increase of the conductivity $\gamma$ of the solution.
Because $\gamma \propto c_\mathrm{i}$ and $\lambda_{\rm{D}} \propto 1/\sqrt{c_\mathrm{i}}$ where $c_\mathrm{i}$ is the ion concentration of the solution, we can expect $f_q \propto \sqrt{c_\mathrm{i}}$. Therefore we fitted the experimental data on $f_q$ with $f_q\propto\sqrt{c_\mathrm{NaCl}+c_0}$, where $c_\mathrm{NaCl}$ is the concentration of NaCl and $c_0$ represents residual ionic contribution in deionized water.

The reason for dissociation of chains below $f_q$ seems to be the misalignment of orientation of Janus particles within the same chain (see Supplemental Movie 3). Fluctuations in the orientation of the Janus particle at the head of the chain cause a relative velocity between the head and the rest of the chain which overcome the attracting force between quadrupoles of adjacent particles. Since the self-propelling velocity is smaller for frequencies higher than $f_v$ [\figref{fig2}(a)], chains are more stable for the fluctuations in the orientation of the Janus particles. On the contrary at low frequency, the self-propelling force is stronger compared with screened attractive electrical interaction. This effect causes dissociation of chains at low enough frequencies.

For deionized water or very low concentrations of NaCl, we do not show $f_v$ and $f_q$, because no precise data could be obtained due to a large drift in ion concentration over time. In addition to that, the polarity of the induced dipole on the silica side sensitively depends on the ion concentration in the solution in this regime since the electrical conductivity of the solution $\gamma_l$ and that of silica, which may be contaminated during thermal evaporation process, $\gamma_p$ are comparable. If $\gamma_l < \gamma_p$ holds, the induced dipole on the silica side is parallel to the electric field as in the metal side. In such configuration, two Janus particles are not attractive. But for frequencies higher than $f_M$, the Janus particles can become quadrupoles which attract each other. The frequency $f_q$ at which chains dissociate may be determined by $f_M$ in such cases. However, again the precise determination of $f_q$ or $f_M$ is difficult due to the drift in the ion concentration in solutions for deionized water.  
 
\section{Flagellar dynamics of chains}
\subsection{Experimental results}
\begin{figure}[tbh]
\begin{center}
\includegraphics[width=138truemm]{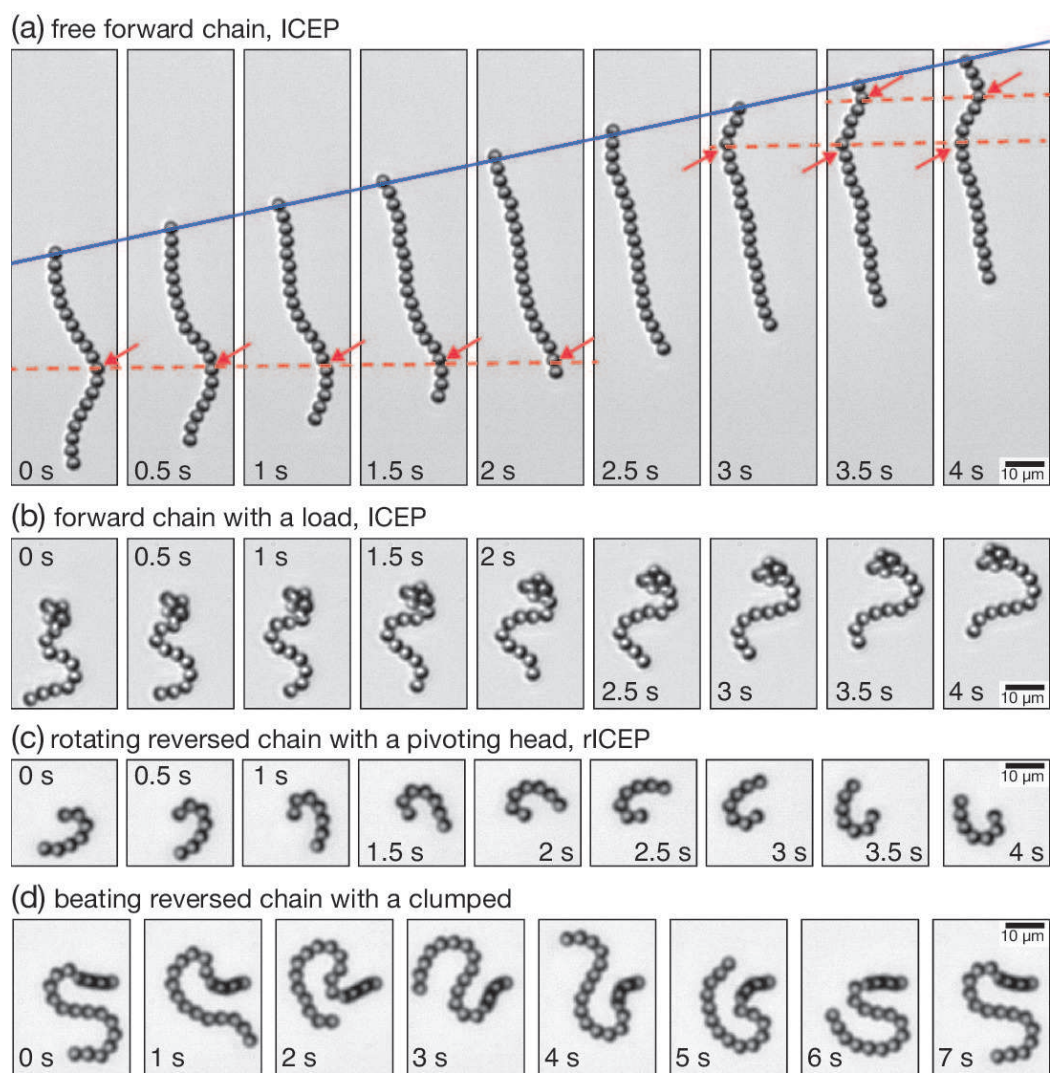}
\caption{
Time-lapse of motions of chains. Intensity is adjusted for visibility.
(a) Wriggling free chain ($V=16$ $\mathrm{V_{pp}}$, 100~kHz, 0.8~mM NaCl, ICEP regime). Blue line is connecting the positions of the heads to visualize almost constant speed of propulsion. Arrows are indicating the peak positions of the wave propagating along the chain. Dashed lines are connecting the peak positions to show that they are almost fixed in the laboratory frame, implying that the phase velocity of the wave is almost equal to the speed to the chain.  
(b) A chain with its head attached to a load can transport the load by exhibiting beating behavior like spermatozoa ($V=20$ $\mathrm{V_{pp}}$, 170~kHz, 1.0mM NaCl, ICEP regime).
(c) A chain with a pivoting head exhibits rotary motion ($V=17$ $\mathrm{V_{pp}}$, 1~MHz, 0.1~mM NaCl, rICEP regime).
(d) Flagellar dynamics of a clamped chain with both positional and orientational constraints ($N=14$, $V=20$ $\mathrm{V_{pp}}$, 1~MHz, 0.1~mM NaCl, rICEP regime).
See also Supplemental Movies 1--7. [Note that the resolution of these figures are lowered for uploading this manuscript to arXiv.]
\label{fig3}}
\end{center}
\end{figure}

A variety of interesting structures are found in the high frequency regime as shown in \figref{fig1}(b), but here we focus on the dynamics of chains. Free chains of Janus particles do self-propel by changing their directions and by showing wiggling motion [\figref{fig3}(a) and Supplemental Movie 2]. When some constraints are imposed on the heads of the chains, they exhibit oscillatory or rotary behavior. A chain with its head attached to an aggregate as a load starts to beat like a flagellum \cite{Isele-Holder2016}, demonstrating its capability of transporting cargos at the microscopic scale [\figref{fig3}(b) and Supplemental Movie 2]. When the fore-most particle of a chain is pivoted, meaning that its position is fixed but its direction can rotate, the chain exhibits rotary motion [\figref{fig3}(c) and Supplemental Movie 4]. An additional constraint on the direction of the fore-most particle of a chain gives rise to stable beating behavior of this clamped chain at sufficiently strong propulsion force [\figref{fig3}(d) and Supplemental Movies 5--7], which was described as a Hopf bifurcation \cite{Sekimoto1995}.

Such constraints can be realized when the fore-most particles happen to be tethered on the bottom electrode by chance or when they hit obstacles or aggregates of immobile particles on the electrodes.
In these cases, because front particles are unable to move forward but are pushed by rear particles, internal stress along the chains accumulates and the straight configuration of the chains becomes unstable, leading to buckling and subsequent rotary or beating behavior quite similar to those of eukaryotic flagella.

Here we investigate the flagellar dynamics of clamped chains of self-propelled Janus particles deep in the rICEP regime at high frequency $f$, because the chains in this regime are more stable than those in the ICEP regime due to relatively smaller self-propelling velocity as we have mentioned in the previous section. Dynamics of 4 chains with the numbers of composing particles $N=$5, 5, 10, 14 were analyzed by changing the applied voltage $V$ between 12 $\mathrm{V_{pp}}$ to 20 $\mathrm{V_{pp}}$ at 1 MHz with 0.1 mM NaCl, where the subscript in $\mathrm{V_{pp}}$ means peak-to-peak voltage (the double amplitude of the sinusoidal wave). The movies were captured at 20~Hz. To make the tracking of each particle easier, the images were intentionally kept slightly off focus. 

\subsection{Time series of configurations: Kymographs and principal components}

\begin{figure}[tbhp]
\begin{center}
\includegraphics[width=138truemm]{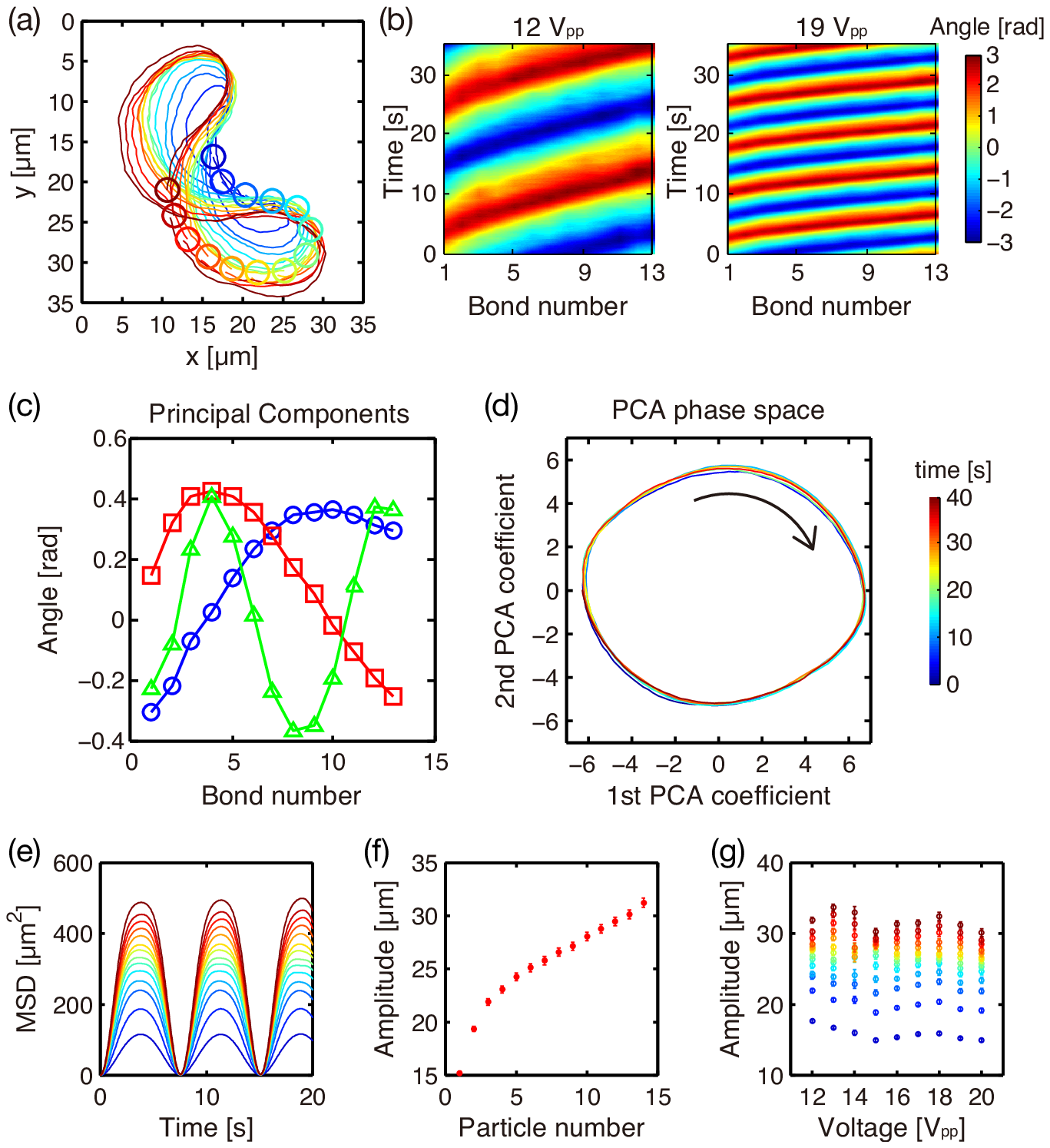}
\caption{
Experimental data on the configurational characteristics of the beating chains. The chain of $N=14$ is shown as an example here. The applied voltage is $V=19$ $\mathrm{V_{pp}}$ except for (b).
(a) Trajectories of all the particles in the chain during 7.5 seconds. See Supplemental Movie 8 for their time evolution.
(b) Kymographs of the bond orientations for $V=12$ $\mathrm{V_{pp}}$ and $V=19$ $\mathrm{V_{pp}}$. The mean values of the angles are adjusted to be 0. The original data with values only on the integer bond numbers were interpolated for visibility. The beating can be recognized as propagating waves.
(c) Principal components obtained by PCA. The 1st mode (blue circles), the 2nd mode (red squares), and the 3rd mode (green triangles) account for 56.4\%, 41.2\%, and 1.16\% of variation respectively. (d) Trajectory in the PCA phase space. The almost overlapping circular trajectory represents a limit cycle.
(e) MSD vs time for all the particles. The oscillatory behavior appears also in the MSD.
(f) Amplitudes of the beating vs the particle number from the front. The amplitudes are estimated from the first peak of MSD shown in (e). Error bars: the standard errors of the estimated MSD at the 1st peak.
(g) Amplitudes vs the applied voltages. Amplitudes stay constant over the change of the voltage.
\label{fig4}}
\end{center}
\end{figure}

To investigate how the shapes of the chains evolve in time, we first tracked all the particles in the chains and then made kymographs of angles of bonds $\phi_i$ ($i=1, 2, \dots, N-1$) connecting the neighboring $i$-th and $(i+1)$-th particles in the chains [\figref{fig4} (a,b)].  As shown in \figref{fig4} (b), the beating of the chains can be regarded as a result of a propagating wave from the head to the tail on the chains. This direction of propagation is different from the prediction by the continuum theory for the case of clamped heads \cite{Camalet1999} but is consistent with experimental observations on swimming spermatozoa \cite{Ishimoto2017} and the theoretical calculation for the case of heads without orientational constraints \cite{Camalet1999}. This suggests that the continuum theories assuming the small amplitudes of oscillations may break down in such large amplitude cases as the beating Janus chains.

Then we applied principal component analysis (PCA) on the time series of the set of bond orientations $\phi_i$ ($i=1, 2, \dots, N-1$) [\figref{fig4}(c,d)].
\figref{fig4}(c) gives an example of the result of the PCA. The 1st and the 2nd modes together explains $97.6\%$ of the variation, while the 3rd mode explains only $1.16\%$. Thus the 1st and the 2nd modes captures almost all the characteristics of the beating behavior. \figref{fig4}(d) depicts the trajectory in the PCA phase space which represents the corresponding limit cycle. It is quite circular like the phase space trajectory of a simple harmonic oscillator, thus the beating behavior can be interpreted as a very simple oscillation between the two PCA modes in spite of the characteristic, even possibly complicated, trajectories of the particles in the real space as shown in \figref{fig4}(a). This simple phase space behavior is in stark contrast to the beating of human sperms \cite{Ishimoto2017} but similar to that of bull sperms \cite{Ma2014}, which may be attributed to their three-dimensional beatings.

To understand how the shapes of the beating behavior depend on the applied voltages,
the amplitudes of beating were evaluated from the mean square displacement (MSD) of each particle in the chains [\figref{fig4}(e,f)].
The MSDs of the particles in the beating chains oscillate, so we used their first peaks as the estimates of the amplitude of oscillations (see \ref{AppMSD} for details). The amplitudes increase rapidly in the first three particles from the heads in the chains, which suggests that these front parts of the chains behave differently from the other rear parts.
This can be interpreted that the buckling, as a result of internal stress accumulation along the chains, occurs in the front regions due to higher stress and that the rear parts are just following their neighboring front particles rather than buckling. In addition, the amplitudes stayed almost constant over the applied voltages [\figref{fig4}(g)], which means that the shapes of the oscillation does not depend on the applied voltages.
These facts may account for the simple behavior of the beating in the PCA phase space. We note that a non-monotonic behavior of amplitudes of beating behavior as a function of the position on a filament was predicted in a continuum model \cite{Camalet1999}.

\subsection{Dispersion relation: Scalings of beating frequency and phase velocity}
\begin{figure}[t]
\begin{center}
\includegraphics[width=138truemm]{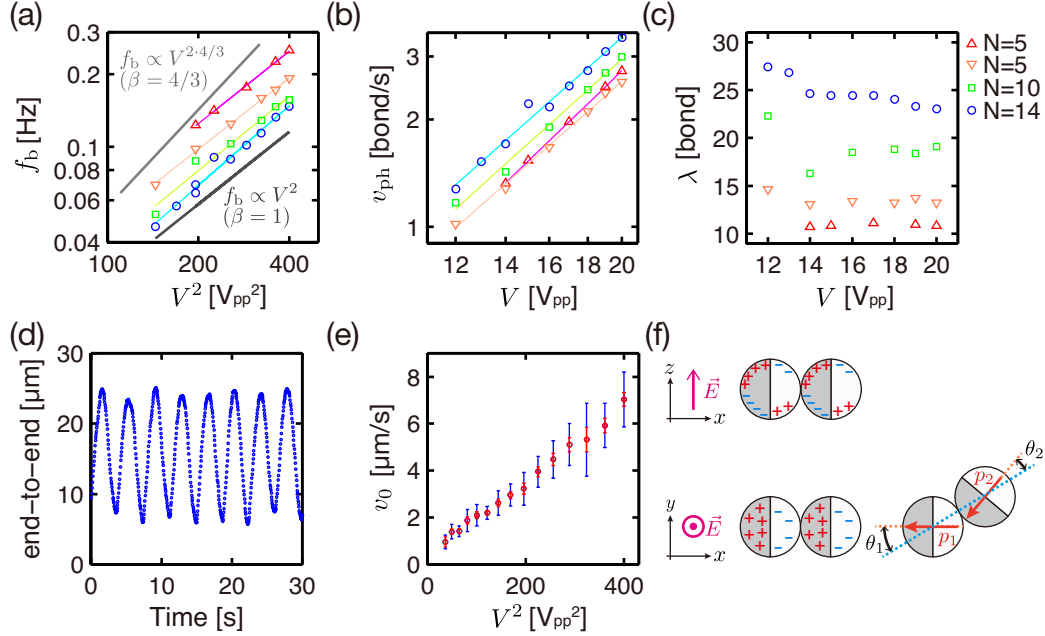}
\caption{
(a-c) Dispersion relations of the beating chains. Estimated exponents are shown in \tabref{table1}.
(a) Beating frequency $f_\mathrm{b}$ vs the applied voltage squared $V^2$ on the log-log scale. Fittings with $f_\mathrm{b}\propto V^{2\beta}$ result in $\beta\simeq 1$ (solid color lines). Solid gray lines corresponds to $\beta=1$ and $\beta=4/3$ to guide the eye.
(b) Phase velocity $v_\mathrm{ph}$ vs the applied voltage $V$ on the log-log scale. Fittings with $\alpha$ $v_{\mathrm{ph}}\propto V^\alpha$ result in $\alpha\simeq 2$ (solid color lines).
(c) Wavelengths $\lambda$ of the wave on the chains as a function of the applied voltage $V$. $\lambda$ was measured in the unit of bond ($\simeq$ diameter). $\lambda$ stays almost constant.
(d) Time series of the end-to-end of the $N=14$ chain at $V=19$ $V_\mathrm{pp}$. Twice the period in this plot corresponds to the actual period of the single beat of the chain.
(e) Self-propulsion speed $v_0$ of the isolated single Janus particles is proportional to the applied voltage squared $V^2$. Shorter red error bars: Standard errors. Longer blue error bars: Standard deviations among the measured particles.
(f) Schematics of the quadrupole-quadrupole interaction between two Janus particles (top) and their effective dipole-dipole interaction in the horizontal plane.
\label{fig5}}
\end{center}
\end{figure}

To understand the scaling behavior and the dispersion of the propagating waves on the chains, we investigated how the beating frequency $f_\mathrm{b}$, the phase velocity $v_\mathrm{ph}$ of the waves, and their wavelength $\lambda$ depend on the applied voltage $V$ [\figref{fig5}(a-c)].

The beating frequency $f_\mathrm{b}$ of the chains becomes higher as the applied voltage increases [\figref{fig5}(a)], where $f_\mathrm{b}$ was extracted from the time series of the end-to-end distance of beating chains by calculating the duration of about 7--15 periods [\figref{fig5}(d)].
This can be ascribed to the faster internal stress accumulation due to the larger propulsive force $F_\mathrm{p}$ at high voltages. 
Both the Langevin simulation on chains of active Brownian particles \cite{Chelakkot2014} and the continuum theory on active flexible filaments \cite{Sekimoto1995,Bourdieu1995} predicted the same scaling relation $f_\mathrm{b}\propto {F_\mathrm{p}}^{\beta}$ with $\beta=4/3$ between the beating frequency $f_\mathrm{b}$ and the self-propulsive force $F_\mathrm{p}$. Because the self-propulsion speed $v_0$ and thus the propulsive force $F_\mathrm{p}$ of the Janus particles are experimentally known to be $F_\mathrm{p}\propto V^2$ \cite{JiangUnpublished,Suzuki2011,Yan2016} and is also confirmed in our experiment [\figref{fig5}(e)], the beating frequency $f_\mathrm{b}$ can be represented with the exponent $\beta$ as $f_\mathrm{b}\propto V^{2\beta}$.

As shown in \figref{fig5}(a) and in \tabref{table1}, the beating frequency $f_\mathrm{b}$ is almost proportional to $V^2$. To be more precise, under the assumption that $\beta$ is constant against $N$ and by taking the overlapping region of the 95\% confidence levels of the estimated $\beta$ for each chain, $\beta$ is estimated to be probably in the range [0.9681, 1.0358] and thus is sufficiently close to $\beta=1$.

Interestingly, the exponent $\beta=1$ obtained in our experiment deviates from the theoretically predicted value $\beta=4/3$ \cite{Bourdieu1995,Chelakkot2014}. However, this discrepancy can be resolved by considering effective dipole-dipole interactions between Janus particles. In the previous work \cite{Chelakkot2014}, dimensional analysis considering energy input due to self-propulsion and viscous dissipation led to,
\begin{equation}
\label{EqBeatingFrequency}
f_\mathrm{b} \sim {F_\mathrm{p}}^{4/3} \zeta^{-1} \kappa^{-1/3},
\end{equation}
where $\kappa$ is the bending rigidity of the chains and $\zeta$ is the friction coefficient of the chain. In our experiment, $F_\mathrm{p}$ of the Janus particles is regulated by the applied voltage $V$ through ICEO flow resulting from the induced charge on the particles. In the theoretical studies \cite{Bourdieu1995,Chelakkot2014}, the bending rigidity $\kappa$ was assumed to be a constant parameter independent of propulsive forces. In contrast, the bending rigidity $\kappa$ of the Janus chains is dependent on the applied voltage $V$ because the attractive force between the particles are also mediated by the electrostatic interactions of the induced charges on the particles.

By assuming that the attractive forces between the Janus particles at high frequency originate from the induced electric quadrupoles on the particles as we discussed in \secref{SecFormationOfChains}, looking from above, the interaction in the horizontal plane can be effectively interpreted as electric dipole-dipole interaction,
${\displaystyle U=\frac{1}{4\pi \epsilon r^3} (\bm{p}_1 \cdot \bm{p}_2 - 3 (\bm{p}_1 \cdot \hat{\bm{r}})(\bm{p}_2 \cdot \hat{\bm{r}})) }$, where $U$ is the interaction potential, $\bm{p}_1$ and $\bm{p}_2$ are dipole moments of the particles, $r$ is the distance between the two dipoles, $\hat{\bm{r}}$ is its unit vector, and $\epsilon$ is the electric permittivity [\figref{fig5}(f)]. Two electric dipoles with the strengths $p_1=|\bm{p}_1|$ and $p_2=|\bm{p}_2|$ placed in 2D as in \figref{fig5}(f) with angular displacements $\theta_1$ and $\theta_2$ from the line connecting their positions exerts a torque $T$ on the dipole $p_2$ as,
\begin{equation}
T=-\frac{\partial U}{\partial \theta_2}=
-\frac{p_1 p_2}{4\pi\epsilon a^3}(\sin{\theta_1}\cos{\theta_2}+2\cos{\theta_1}\sin{\theta_2}) \propto -\frac{p_1 p_2}{2\pi\epsilon a^3} \theta_2 \propto -V^2 \theta_2,
\end{equation}
where we assumed that the bending displacement is small, $\theta_{1,2} \ll 1$, and left the lowest $\theta_2$ term. The last proportionality comes from the fact that the strength of the induced dipole is proportional to the applied voltage. This torque $T$ works as a restoration force to decrease the deviation from the straight configuration. This is responsible for the bending rigidity $\kappa$ of the chain. Therefore, the above argument gives the following dependence,
\begin{equation}
\label{EqKappaV}
\kappa\propto T \propto V^2.
\end{equation}
Finally, by inserting both \equref{EqKappaV} and $F_\mathrm{p}\propto V^2$ into \equref{EqBeatingFrequency}, because the friction coefficient $\zeta$ does not depend on $V$, we obtain,
\begin{equation}
f_\mathrm{b}\propto {(V^2)}^{4/3} (V^2)^{-1/3} \propto V^2 \propto {F_\mathrm{p}}^1.
\end{equation}
In terms of $\beta$ defined as $f_\mathrm{b}\propto {F_\mathrm{p}}^{\beta}$, consideration of the voltage dependence of the bending rigidity $\kappa$ leads to $\beta=1$, which is consistent with our experimental result.

The above argument and our experimental result $\beta=1$ support the idea that electric quadrupoles are induced on Janus particles at the high frequency regime and that they are responsible for the attractive interaction between them. Although the difficulty of the direct measurement of the induced charges on the particles has hindered the understanding of the precise mechanisms of the Janus particles, our measurement gives the first experimental evidence of the surface charge distribution on the Janus particles.

\begin{table}
\caption{\label{table}Scaling exponents estimated from the experimental data. Errors ($\pm$) mean standard errors of linear fitting. 95\% confidence levels are shown in the following brackets.
\label{table1}
}
\begin{indented}
\item[]\begin{tabular}{@{}llll}
\br
$N$ & $\alpha$ ($v_{\mathrm{ph}}\propto V^\alpha$) & $\beta$ ($f_\mathrm{b}\propto V^{2\beta}$) \\
\mr
$5$ &$2.028\pm 0.015$, [1.9808, 2.0747] & $0.998\pm 0.022$, [0.9277, 1.0691] \\
$5$ &$1.704\pm 0.054$, [1.7035, 2.0016] & $0.985\pm 0.018$, [0.9348, 1.0358] \\
$10$ &$1.894\pm 0.079$, [1.6757, 2.1127] & $1.026\pm 0.083$, [0.7967, 1.2559] \\
$14$ &$1.864\pm 0.093$, [1.6454, 2.0828] & $1.098\pm 0.056$, [0.9681, 1.2284] \\
\br
\end{tabular}
\end{indented}
\end{table}

We also calculated the phase velocity $v_\mathrm{ph}$ of the waves on the chains and its scaling relation $v_\mathrm{ph}\propto V^\alpha$ [\figref{fig5}(b)]. The phase velocities were estimated by finding the direction with the smallest variation of the kymographs. This was done by the structure tensor method \cite{Rezakhaniha2012,Nishiguchi2017,Kawaguchi2017}.
Linear fitting on the log-log scale gives the exponent $\alpha$ close to 2. Again by taking the overlaps of the 95\% confidence levels, the estimate of $\alpha$ resides in [1.9808, 2.0016], thus it is probable that $\alpha=2$.

The two experimentally extracted exponents $\alpha$ and $\beta$ means the wavelengths of the waves on the chains are constant. By defining the scaling exponent of the wavelength $\lambda$ as $\lambda\propto V^\gamma$, the relation $v_\mathrm{ph}=f_\mathrm{b}\lambda$ gives a hyperscaling relation $\alpha=2\beta+\gamma$. The estimated values $\alpha\simeq 2$ and $\beta\simeq 1$ mean $\gamma \simeq 0$. As a matter of fact, the wavelengths $\lambda$ estimated experimentally by using $v_\mathrm{ph}=f_\mathrm{b}\lambda$ can be regarded to be constant over the whole range of the applied voltages [\figref{fig5}(c)], although the data are quite noisy.  
This as well as the constant oscillation amplitudes discussed in \figref{fig4}(g) indicates that the shapes and the trajectories of the beating are all the same but only the timescale changes according to the applied voltage. 
Interestingly, the wavelength $\lambda$ increases as the increase of the length of the chain $N$. This may also signify that only a couple of particles in the front part play important roles and the other rear particles are just following their front neighbors.

\subsection{Effect of hydrodynamics}
\label{SecHydro}
Finally, we theoretically estimated the flow field created around the chains in order to examine their role in the flagellar dynamics.
The flow field created by Janus particles is quite complicated because of the broken symmetries. Not only the distinct hemispheres of Janus particles but also the applied vertical electric field breaks the symmetry of the flow, so it is not even axisymmetric. However, as it was analytically calculated for the low frequency ICEP regime \cite{Nishiguchi2015} and was also experimentally measured for a Janus particle but with different configurations of walls and electrodes as ours \cite{Peng2014}, there exists a pusher-type dipolar flow at the lowest order. Although the existence of the bottom electrode very close to the particles further complicates the actual flow field, we calculated the flow field around the chains by assuming that pusher-type dipolar flows are created around each particle as a first approximation.

\begin{figure}[t]
\begin{center}
\includegraphics[width=115truemm]{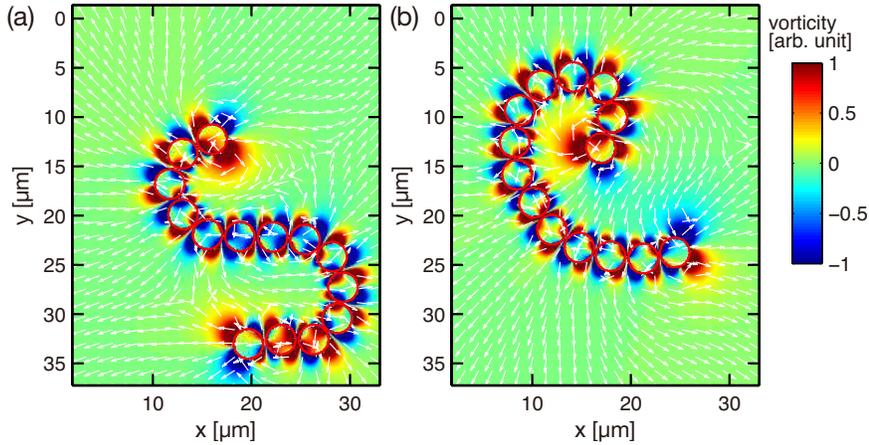}
\caption{
Flow field calculated by superposing the pusher-type dipolar flow at the centers of the experimentally obtained particle positions at two different configurations. Directions of the flow is depicted by white vectors. The background color corresponds to the vorticity in arbitrary unit. See Supplemental Movie 8 for the time series of the flow field.
\label{fig6}}
\end{center}
\end{figure}

The whole flow field can be obtained by superposing the dipolar flows from all the particles in the chain (see \ref{AppFlowField} for details). Because precise detection of particle orientations $\psi_i$ from the images is a difficult task, we approximated $\psi_i$ so that the bond orientation $\phi_i$ satisfies $\phi_i=(\psi_i+\psi_{i+1})/2$ ($i=1, 2, \dots, N-1$). By arranging the force dipoles directing $\psi_i$ at the experimentally detected particle positions, we obtained the flow field as shown in \figref{fig6} and Supplemental Movie 9.

Hydrodynamic couplings between the Janus particles composing the chains work to enhance buckling and thus beating behavior. From the calculated flow field, the surrounding fluid flows inward at the concave side but outward at the convex side. These flows apply distorting forces on the chain and further facilitates the bending and the beating behavior. A similar effect was also discussed previously that such hydrodynamic instability can cause beating behaviors \cite{Jayaraman2012,Laskar2015}.

This possibility of spontaneous beating due to the hydrodynamic instability even in the case of free chains without any loads on their heads raises an interesting question that we may explore experimentally: Does this beating enhance the speed of chains or not?
In this regard, it is noteworthy that there exist weak beating-like behaviors even in almost-straight chains (see \figref{fig2}(a)) and large-amplitude beating in chains with loads on their heads (see \figref{fig2}(b)). In both cases, the heads of chains are not anchored and the whole chain are moving forward accompanying the beating behaviors. 
We will discuss this issue in the next subsection.    

\subsection{``Beating by propulsion'' vs ``Propulsion by beating''}
Up to here we have discussed beating of chains of Janus particles in analogy with flagellar dynamics in biological systems.
In this subsection, we discuss potential differences between ``propulsion by beating'' and ``beating by propulsion''. Eukaryotic flagella or cilia enable self-propulsion by producing propagating waves (beating in two dimension or helical wave in three dimension) at low Reynolds number. This type of dynamics, to name ``propulsion by beating'', enables swimming only by creating waves. On other hand, chains of active particles presented in this paper and actin filaments on myosin beds \cite{Sekimoto1995, Bourdieu1995} can propel even without beating, because active forces are acting uniformly on the chains or the filaments in these examples. Beating behaviors in these cases are produced due to buckling by propulsive force or the enhancement by hydrodynamic interaction \cite{Jayaraman2012,Laskar2015}. We call the beating produced in this kind of situation ``beating by propulsion''.

How the beating affects swimming speed can be different in two cases. As was calculated by G.I. Taylor \cite{Taylor1951,Lauga2009}, swimming speed $U$ of a sheet producing a transversal wave in viscous fluid is given by $U = (1/2)\omega k b^2$ for a small amplitude ${b}$ of the oscillation,  the angular frequency $\omega$ and the wave number $k$. Here, the sheet is a simplified representation of a flagellum in two dimension.  Although the generalization to the beating thin cylinder were performed, the essence was the same \cite{Taylor1952}. Thanks to the symmetry of $b \rightarrow -b$, $U$ is proportional to $b^2$, and thus there is literally no swimming without beating in the case of ``propulsion by beating''.. 
In contrast, in the ``beating by propulsion'' case, chains or filaments can move forward even without beating. How the swimming speed of filaments exhibiting ``beating by propulsion'' depends on the beating parameters is an interesting question. One can superimpose pusher-type flows in the same argument as the sheet or the cylinder producing transversal waves. It gives the relation $U = U_0 + C \omega k b^2$, where $U_0$ is the bare propulsion speed of active sheets and $C$ is a constant. In the cylinder approximation, terms expanded by the series of $ak$ with $a$ being the diameter will be added \cite{Taylor1952}.

The result implies that, when the beating is spontaneously created in the filaments, the swimming speed will be enhanced due to the beating. In the case of actin filaments on myosin beds, hydrodynamic effect must be very weak, and thus the enhancement of the propulsion speed by beating cannot be expected. While in chains of active particles, spontaneous beating and subsequent enhancement of the propulsion speed can be triggered due to stronger hydrodynamic interaction. To test whether this hydrodynamic instability actually increases the propulsion speed or not is a future experimental challenge.

\section{Conclusion and outlook}
Our experimental investigation on the behavior of self-propelling Janus particles by varying salt concentration and the frequency of the applied voltage has revealed that there exist two characteristic frequencies for the velocity reversal and for the interaction switching. Especially, we found a new phase in which the Janus particles swimming toward the dielectric side (usual ICEP) can also show attractive interactions under sufficiently high ion concentration.
This signifies that the propulsion mechanism and the interaction mechanism have to be considered separately.

The experimentally obtained scaling relations of the flagellar dynamics of the chains composed of the Janus particles demonstrate that the modification of the applied voltage has only the effect on the timescales of the dynamics. The phase velocity $v_\mathrm{ph}$ and the beating frequency $f_\mathrm{b}$ are experimentally verified to scale as $V^2$, which is the same dependence as the self-propulsion force $F_\mathrm{p}$. This dependence could be explained by modifying the scaling theory \cite{Chelakkot2014} by assuming electric quadrupoles on the particles, which gave the indirect but convincing evidence for the charge distribution on the Janus particles.

Because such $V^2$ dependence originates from the fact that both the self-propulsion force arising from ICEO and the strength of the electrostatic interactions between the induced charges have exactly the same scaling $\propto V^2$,
different interaction mechanisms would result in different scalings for $v_\mathrm{ph}$ and $f_\mathrm{b}$.
Therefore, fine-tuning this dependence may be crucial for realizing beating behavior optimal for biological functions which will lead to providing a new design principle for engineered microswimmers.

Our calculation on the flow field around the chain is still a quite rough estimation. However, the obtained flow field captures a profile similar to that of eukaryotic flagella \cite{Ishimoto2017}. This can be a characteristic of the flow field around active filaments.
For more precise understanding, confocal imaging of the flow field is left for future work. Experimental observation of the actual flow field would enable direct comparison of the beating of the Janus chains and that of eukaryotic flagella \cite{Ishimoto2017}.
Further investigations on the hydrodynamic effect of the beating on the swimming speed, especially the dependence on the parameters such as the beating amplitude, the beating frequency, the wave number and the length of chains, will make it much clearer whether the beating of the chains of the Janus particles is triggered by mechanical buckling or by hydrodynamic instability and lead to better understanding of swimming strategy in the microbial world. 

\section*{Acknowledgements}
This work was supported by a Grant-in-Aid for the Japan Society for Promotion of Science (JSPS) Fellows (Grant No. 26-9915), KAKENHI (Grants No. 25103004, ``Fluctuation \& Structure'') from MEXT, Japan,

\appendix
\section{Estimation of the self-propelling speed and the beating amplitude from MSD}
\label{AppMSD}
\subsection{Self-propelling velocity $v_0$}
The MSD of a self-propelled particle with velocity $v_0$ subject to rotational and translational Brownian noise in two dimension is given by \cite{Jiang2010},
\begin{equation}
\label{EqMSD}
\mathrm{MSD}(t)=\left(4D+\frac{2 {v_0}^2}{D_\mathrm{r}}\right) t+\frac{2 {v_0}^2}{{D_\mathrm{r}}^2}\left( \mathrm{e}^{-D_\mathrm{r} t}-1 \right),
\end{equation}
where $D$ and $D_\mathrm{r}$ are translational and rotational diffusion coefficients respectively. Fitting the experimentally obtained MSDs with this nonlinear function is not so accurate when we do not have sufficiently long trajectories, so we used its short-time approximation. In the time scale where the rotational diffusion can be neglected ($t\ll 1/D_{\mathrm{r}}$), \equref{EqMSD} can be rewritten as,
\begin{equation}
\label{EqMSDshorttime}
\mathrm{MSD}(t)\approx 4Dt+{v_0}^2 t^2.
\end{equation}
Although the existence of the wall very close to the particle and nonequilibrium fluctuations in our experiment require some correction, we can estimate $D_{\mathrm{r}}$ from the Einstein-Stokes relation.
In our experiment with the particle diameter $d=3.17$ $\mathrm{\mu m}$, the Einstein-Stokes relation gives $1/D_{\mathrm{r}}=\pi \eta d^3/k_B T=24.2$ s, where $\eta$ is the viscosity of the surrounding fluid, $k_B$ is the Boltzmann constant, and $T$ is the absolute temperature. Therefore, we used experimentally obtained MSDs upto 1 or 2 seconds and fitted them with \equref{EqMSDshorttime} to obtain $v_0$.

We note that this gives an accurate estimate of $v_0$ compared with just taking the difference of the time series especially when $v_0$ is small.

\subsection{Beating amplitudes}
We estimated the amplitudes of the beating behavior from the first peak of the oscillatory MSDs as shown in \figref{fig4}(e) by,
\begin{equation}
\label{EqAmplitude}
(\mbox{Amplitude})=\sqrt{2\times\mbox{(1st peak of MSD)}}.
\end{equation}
We chose this as an estimate because the double amplitude of a simple harmonic oscillator can be estimated by \equref{EqAmplitude}.

For simplicity, we consider a one-dimensional oscillator.
Consider a harmonic oscillator with the position $x(t)$ given by,
\begin{equation}
x(t)=A\sin{\omega t},
\end{equation}
where $A$ is the amplitude and $\omega$ is the angular frequency.
The MSD as a function of delay time $\tau$ is given by,
\begin{eqnarray}
\mathrm{MSD}(\tau)&=\langle [x(t+\tau)-x(t)]^2 \rangle_t \\
&=\left\langle 4A\sin^2{\frac{\omega}{2} \tau}\cos^2{\left(\omega t + \frac{\omega}{2}\tau\right)}  \right\rangle_t  \\
&=2A^2\sin^2\frac{\omega}{2}\tau,
\end{eqnarray}
where $\langle \; \rangle_t$ is the average over $t$. Thus, the amplitude $A$ is estimated from the MSD as,
\begin{equation}
A=\sqrt{\frac{1}{2}\times\mbox{(1st peak of MSD)}}.
\end{equation}
Therefore, we can estimate the double amplitude, or the peak-to-peak amplitude, as \equref{EqAmplitude}.

\section{Calculation of the flow field}
\label{AppFlowField}
Here we detail the calculation procedure to obtain the flow field presented in \secref{SecHydro}.

Pusher-type dipolar flow $\bm{u}(\bm{r},\bm{r}_0,\bm{p})$ at position $\bm{r}$ created by a single force dipole $\bm{p}$ at $\bm{r}_0$ is mathematically represented as \cite{Lauga2009},
\begin{equation}
\label{EqDipolarFlow}
\bm{u}(\bm{r},\bm{r}_0,\bm{p})=\frac{p}{8\pi\eta |\bm{r}-\bm{r}_0|^2}[3\cos^2{\Theta}-1]\frac{\bm{r}-\bm{r}_0}{|\bm{r}-\bm{r}_0|},
\end{equation}
where $p$ is the signed strength of the dipole with $p>0$ for pushers and $p<0$ for pullers, $\eta$ is the viscosity of the fluid, and $\Theta$ is the angle between the orientation of the force dipole $\bm{p}$ and the relative position vector $\bm{r}-\bm{r}_0$. In practice, we used the relation,
\begin{equation}
\cos{\Theta}=\frac{\bm{p}\cdot(\bm{r}-\bm{r}_0)|}{\bm{p}\cdot(\bm{r}-\bm{r}_0)|}.
\end{equation}

To calculate the flow field created by pusher-type force dipoles arranged at the experimentally detected particle positions in a chain composed of $N$ particles, we need the set of all the particle positions and the orientations of the dipoles. Here we denote the position and the orientation of the $i$-th particle as $\bm{r}_i$ and $\psi_i$ respectively. We used experimentally obtained centers of mass of the particles as $\bm{r}_i$. Because detecting the polarities of the particles is quite difficult, we estimated $\phi_i$ from the bond orientation $\phi_i$, which is the angle of $\bm{r}_i-\bm{r}_{i+1}$. We chose $\psi_i$ for $i=1, 2, \dots, N$ so that they satisfy $\phi_i=(\psi_i+\psi_{i+1})/2$. To be more precise,
\begin{eqnarray}
\psi_i=\frac{\phi_{i-1}+\phi_{i}}{2}\;\;\mbox{for $i$=2, 3, \dots, N-1},\\
\psi_1=\frac{3\phi_1+\phi_{2}}{2},\\
\psi_N=\frac{3\phi_{N-1}+\phi_{N-2}}{2}.
\end{eqnarray}
By using this $\psi_i$ as the orientation of the $i$-th dipole $\bm{p}_i$, we can calculate the flow field $\bm{u}_i$ created by the $i$-th Janus particle (the $i$-th dipole) from \equref{EqDipolarFlow}. In actual calculation, to avoid the divergence around $\bm{r}\to\bm{r}_i$, we neglected the flow field created by the $\i$-th dipole inside the $i$-th Janus particle as,
\begin{eqnarray}
\bm{u}_i=
\left\{
\begin{array}{ll}
    \bm{u}(\bm{r},\bm{r}_i,\bm{p}) &  \mbox{for}\;\; |\bm{r}-\bm{r}_0|\geq d/2\\
    \bm{0} &  \mbox{for}\;\; |\bm{r}-\bm{r}_0|< d/2
    \end{array}
    \right. ,
\end{eqnarray}
where $d$ is the diameter of the Janus particles.

To obtain the total flow field $\bm{u}_\mathrm{tot}$ created by the whole chain, we assumed that the strength of the dipoles are the same for all the particles and superposed the flow field created by each particle as $\bm{u}_\mathrm{tot}=\sum_{i=1}^N \bm{u}_i$. The results are shown in \figref{fig6} and Supplemental Movie 9.

\section{Supplemental movies}
\begin{description}
\item[Supplemental Movie 1] (1\_ReversedChains\_30Vpp\_1MHz\_01mM\_x1speed.mp4)\\
Experimental movie of reversed chains in the rICEP regime. Continuous reconfigurations of chains can be observed. Conditions: 30~$\mathrm{V_{pp}}$, 1~MHz, 0.1~mM NaCl. The field of view is $124\;\mathrm{\mu m}\times 88 \;\mathrm{\mu m}$. The movie is played at the real speed.
See also \figref{fig1}(b).

\item[Supplemental Movie 2] (2\_ForwardChains\_16Vpp\_150kHz\_1mM\_x3speed.mp4)\\
Experimental movie of forward chains in the ICEP regime at high NaCl concentration (1~mM). Conditions: 16~$\mathrm{V_{pp}}$, 150~kHz, 1.0~mM NaCl. The field of view is $410\;\mathrm{\mu m}\times 328 \;\mathrm{\mu m}$. The movie is 3 times accelerated. Original images of $3000 \times 2400$ pixels are resized to $1344\times 1072$ pixels for reducing the file size.

\item[Supplemental Movie 3] (3\_ChainDissociation\_16Vpp\_20kHz-10kHz\_03mM\_x1speed.mp4)\\
Experimental movie of dissociating forward chains in the ICEP regime. Dissociation of the chains can be observed as the AC field frequency is lowered from 20 kHz to 10 kHz. Conditions: 16 V$_{\mathrm{pp}}$, 20 $\rightarrow$ 10 kHz, 0.3 mM NaCl. The field of view is $82\;\mathrm{\mu m}\times74\;\mathrm{\mu m}$. The movie is played at the real speed. .

\item[Supplemental Movie 4] (4\_RotatingChain\_17Vpp\_1MHz\_01mM\_x1speed.mp4)\\
Experimental movie of a tethered reversed chain in the rICEP regime. The chain rotates around the pivoting fore-most particle. Conditions: 17~$\mathrm{V_{pp}}$, 1~MHz, 0.1~mM NaCl. The field of view is $24.5\;\mathrm{\mu m}\times 24.5 \;\mathrm{\mu m}$. The movie is played at the real speed.
See also \figref{fig3}(c).

\item[Supplemental Movie 5] (5\_BeatingChain-N5\_14-17-20Vpp\_1MHz\_01mM\_x1speed.mp4)\\
Experimental movies of a clamped reversed chain in the rICEP regime with the number of composing particles $N=5$. The beating behavior is observed and its beating frequency $f_\mathrm{b}$ becomes larger as the applied voltage increases. Three movies of the chain with different voltages are aligned horizontally (14~$\mathrm{V_{pp}}$, 17~$\mathrm{V_{pp}}$, 20~$\mathrm{V_{pp}}$). The field of view of each image is $23.6\;\mathrm{\mu m}\times 24.7 \;\mathrm{\mu m}$. Conditions:  1~MHz, 0.1~mM NaCl. The movie is played at the real speed.

\item[Supplemental Movie 6] (6\_BeatingChain-N10\_12-16-20Vpp\_1MHz\_01mM\_x1speed.mp4)\\
Experimental movies of a clamped reversed chain in the rICEP regime with $N=10$. Three movies of the chain with different voltages are aligned horizontally (12~$\mathrm{V_{pp}}$, 16~$\mathrm{V_{pp}}$, 20~$\mathrm{V_{pp}}$). The field of view of each image is $33.0\;\mathrm{\mu m}\times 34.3 \;\mathrm{\mu m}$. Conditions:  1~MHz, 0.1~mM NaCl. The movie is played at the real speed.

\item[Supplemental Movie 7] (7\_BeatingChain-N14\_13-17-20Vpp\_1MHz\_01mM\_x1speed.mp4)\\
Experimental movies of a clamped reversed chain in the rICEP regime with $N=14$. Three movies of the chain with different voltages are aligned horizontally (13~$\mathrm{V_{pp}}$, 17~$\mathrm{V_{pp}}$, 20~$\mathrm{V_{pp}}$). The field of view of each image is $34.6\;\mathrm{\mu m}\times 44.0 \;\mathrm{\mu m}$. Conditions:  1~MHz, 0.1~mM NaCl. The movie is played at the real speed. See also \figref{fig3}(d).

\item[Supplemental Movie 8] (8\_BeatingTrajectory\_N14\_19Vpp\_1MHz\_01mM\_x1speed.mp4)\\
Trajectories of all the particles in a beating chain of $N=14$ during 7.5 seconds. The chain is the same as the Supplemental Movie 6. Conditions: 19~$\mathrm{V_{pp}}$, 1~MHz, 0.1~mM NaCl. The movie is played at the real speed.
See also \figref{fig4}(a).

\item[Supplemental Movie 9] (9\_FlowField\_N14\_19Vpp\_1MHz\_01mM\_x1speed.mp4)\\
Time series of the flow field calculated by using the experimentally obtained particle positions and by assuming the dipolar flow around each particle. Experimental data for the $N=14$ chain at $V=19$~$\mathrm{V_{pp}}$, 1~MHz, 0.1~mM NaCl are used. Arrows represent the local directions of the flow field. The color represents vorticity in an arbitrary unit. The movie is played at the real speed. See also \figref{fig6} and \ref{AppFlowField}.
\end{description}

\bibliographystyle{unsrt}
\bibliography{ref_chain7}

\end{document}